\documentclass[prb,aps,superscriptaddress,reprint]{revtex4-2}
\usepackage{graphicx}
\usepackage{amsmath}
\usepackage{bbold}
\usepackage{amsbsy}
\usepackage{float}
\usepackage{amssymb}
\usepackage[makeroom]{cancel}
\usepackage[version=4]{mhchem}
\usepackage[english]{babel}
\usepackage{xcolor}
\usepackage{bbm}
\usepackage{bm}
\usepackage{siunitx}

\usepackage[normalem]{ulem}

\usepackage{hyperref}

\begin{document}
\title{SQUID pattern disruption in transition metal dichalcogenide\\ Josephson junctions due to non-parabolic dispersion of the edge states}

\author{D. Sticlet}
\affiliation{National Institute for Research and Development of Isotopic and Molecular Technologies, 67-103 Donat, 400293 Cluj-Napoca, Romania}

\author{P. W{\'o}jcik}
\affiliation{AGH University of Science and Technology, Faculty of Physics and Applied Computer Science, al. A. Mickiewicza 30, 30-059 Krakow, Poland}

\author{M. P. Nowak}\email{Corresponding author: mpnowak@agh.edu.pl}
\affiliation{AGH University of Science and Technology, Academic Centre for Materials and Nanotechnology, al. A. Mickiewicza 30, 30-059 Krakow, Poland}

\date{\today}

\begin{abstract}
We theoretically study Josephson junctions with a transition metal dichalcogenide zigzag ribbon as a weak link. We demonstrate that the spatial profile of the supercurrent carried by the edge modes determines the critical current dependence on the perpendicular magnetic field. We explore this finding and analyze the impact of Zeeman interaction and the orbital effects of the magnetic field on the Andreev bound states energies. We show that the unequal Fermi velocities of the spin-opposite edge modes lead to an anomalous shift of the Andreev bound states in the presence of the magnetic field. This is manifested in a pronounced modification of the SQUID critical current oscillations when two opposite edges of the ribbon are conducting and can be exploited in order to reveal the anomalous phase shift of the Andreev bound states in a single Josephson junction device.
\end{abstract}

\maketitle

\section{Introduction}
In a Josephson junction created by linking two superconductors with a piece of normal (e.g.,~semiconducting) material supercurrent can be carried over considerable distance by Andreev bound states (ABS)~\cite{beenakker_universal_1991}. Electrical tunability of the transport properties of the normal part allows us to tailor the supercurrent~\cite{doh_tunable_2005} and its spatial distribution as demonstrated by adjusting the Fraunhofer~\cite{amado_electrostatic_2013} or SQUID~\cite{guiducci_full_2019} interference patterns in 2DEG or graphene-based~\cite{calado_ballistic_2015, ben_shalom_quantum_2016, kraft_tailoring_2018} superconductor-normal-superconductor (SNS) junctions. 

Probing the maximal supercurrent carried through the junction---the critical current---in an external magnetic field enables us to determine the supercurrent density profile and consequently reveals the nature of the transport in the weak link~\cite{dynes_supercurrent_1971}. 
This is exploited in the search of a combination of superconductivity with quantum Hall~\cite{van_ostaay_spin-triplet_2011, amet_supercurrent_2016, lee_inducing_2017} or spin Hall phases~\cite{hart_induced_2014, bocquillon_gapless_2017} as well as for distinguishing~\cite{galambos_superconducting_2020} the topologically protected~\cite{pribiag_edge-mode_2015, blasi_manipulation_2019} from trivial edge states as present in 2DEG SNS junctions~\cite{de_vries_he_2018}. The study of the critical current in SNS junctions realized on atom-thick, layered materials becomes of particular importance for unveiling the conducting edge modes present due to specific atomic edge termination, as demonstrated recently for $\ce{Bi2O2Se}$~\cite{ying_magnitude_2020}.

In this paper we investigate properties of Josephson junctions realized on a newly emerging class of two-dimensional (2D) semiconductors---transition metal dichalcogenides (TMDCs)---serving as a weak link between two superconductors. TMDCs can be tailored into narrow single-layer ribbons\cite{li_vapourliquidsolid_2018, yang_deriving_2019, kotekar-patil_coulomb_2019} and already have been used to create gated structures as field-effect transistors~\cite{radisavljevic_single-layer_2011} and quantum point contacts \cite{marinov_resolving_2017,sharma_split-gated_2017}. 
As predicted by density functional theory calculations~\cite{bollinger_one-dimensional_2001, bollinger_atomic_2003, li_mos2_2008, erdogan_transport_2012}, tight-binding~\cite{rostami_edge_2016}, and continuum~\cite{peterfalvi_boundary_2015} modeling, zigzag ribbons conduct through the edge states in the energy gap of the bulk material. The presence of the edge modes can be visualized by spatial current mapping~\cite{wu_uncovering_2016, prokop_scanning_2020} or scanning tunneling microscopy measurements~\cite{zhang_direct_2014, koos_stm_2016}.

Here we show that in a TMDC SNS junction the critical current dependence on the external magnetic field reflects the number of occupied edges. Most importantly, the critical current patterns reveal unusual dispersion at the edges, which induces an anomalous shift of the ABS in the presence of the magnetic field.

The anomalous ABS structure with $E_i(\phi)\neq E_i(-\phi)$ (where $E_i$ are the ABS energies and $\phi$ is the superconducting phase difference between the leads), is obtained when both time-reversal and chiral symmetries are broken~\cite{krive_chiral_2004}. The first one is violated due to the presence of the magnetic field. Breaking of the second one---the symmetry of leftward and rightward transport process in each spin band---has been predicted as due to the combined effects of band mixing and strong Rashba spin-orbit (SO) coupling in multimode nanowires~\cite{yokoyama_josephson_2013, yokoyama_anomalous_2014, campagnano_spinorbit_2015} or in quantum dots~\cite{dellanna_josephson_2007, zazunov_anomalous_2009, brunetti_anomalous_2013}.
Here we show that this effect appears inherently in TMDC nanoribbons as a result of strongly non-parabolic edge bands and intrinsic SO coupling.

Experimentally the measurement of the anomalous ABS spectrum is realized by combining two Josephson junctions: anomalous and normal one, into a SQUID loop~\cite{szombati_josephson_2016, mayer_gate_2020}.
We show that the anomalous shift can be detected by probing the perturbation of the SQUID pattern in a {\it single} TMDC SNS junction thanks to the simultaneous population of the two edges of the ribbon. Furthermore, we show that in TMDCs the anomalous shift of the ABS is driven not only by Zeeman splitting but also by orbital effects of the perpendicular magnetic field.

This paper is organized as follows. In Sec.~\ref{sec:num_mod} we outline the numerical model. In Sec.~\ref{sec:th_edge} we explain the magnetic field effects on the ABS spectrum and the resulting critical current pattern, focusing on the region where the edge states have an almost parabolic dispersion.
In Sec.~\ref{sec:nonpar} we show how the presence of the non-parabolic bands reveals itself in critical current maps. The conclusions are given in Sec.~\ref{sec:summ}.

\section{Numerical model}
\label{sec:num_mod}

\begin{figure}[ht!]
\includegraphics[width = 8.5cm]{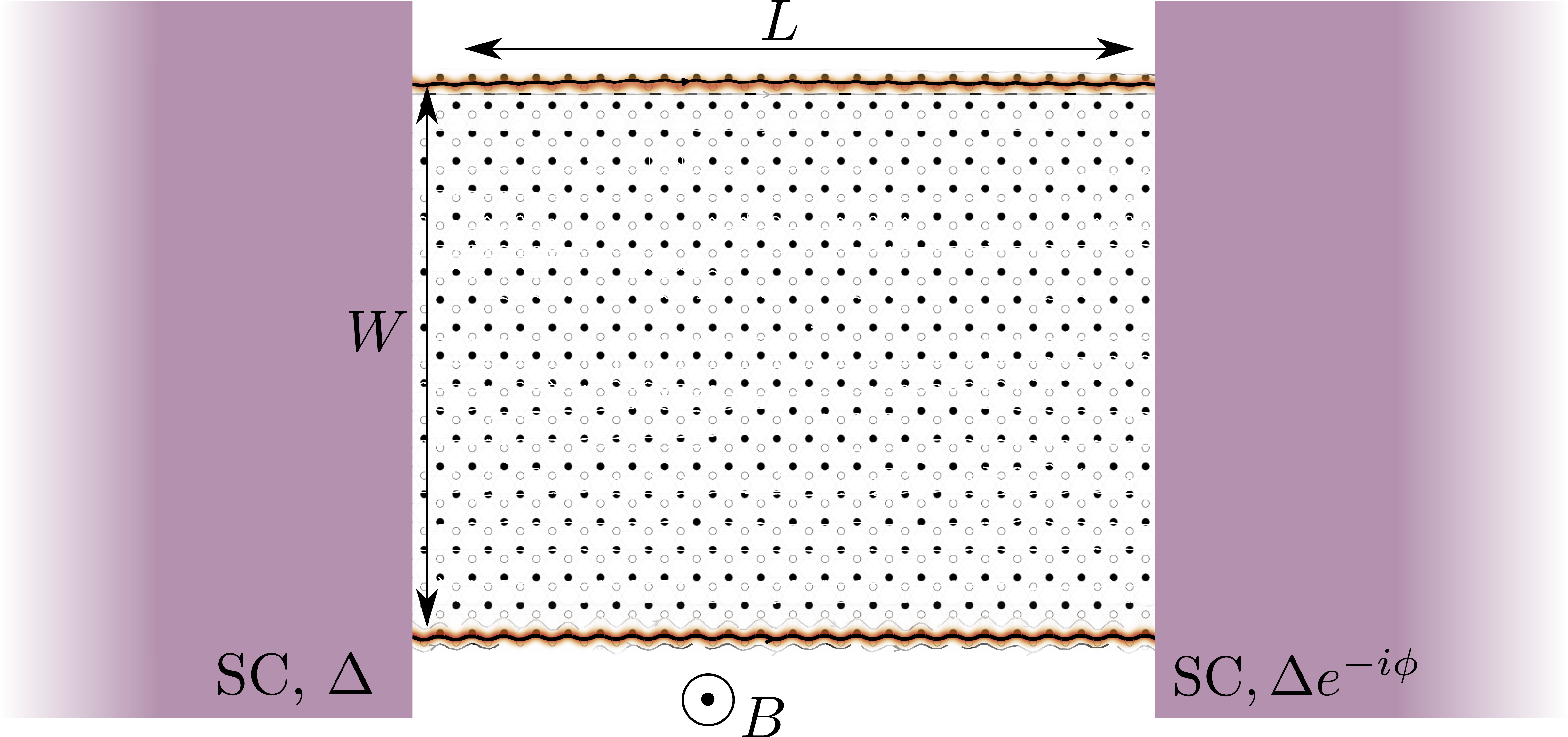}
\caption{Illustration of the considered SNS junction.
A TMDC zigzag nanowire connected with two superconducting electrodes (gray-pink) is threaded by the magnetic flux $\Phi = BWL$. Edge current density is denoted by orange colors.}
\label{system}
\end{figure} 

\subsection{Normal scattering region}
The normal part of the considered SNS junction (Fig.~\ref{system}) is a TMDC \ce{MX2} monolayer shaped into a zigzag nanoribbon. To describe the ribbon we exploit the tight-binding model that contains contributions from $d$ and $p$ orbitals of the metal M and chalcogen X atoms, respectively~\cite{cappelluti_tight-binding_2013, silva-guillen_electronic_2016}. Owing to the system symmetry in the $z$ direction, we perform transformation of the basis that casts the $p$ orbitals of the X layers into symmetric and antisymmetric combinations~\cite{rostami_theory_2015, rostami_valley_2015}.  
The Hilbert space of the final model is spanned by the vector $(d_{3z^2-r^2}, d_{x^2-y^2}, d_{xy}, p_x^S, p_y^S, p_z^A)$, where $S$ and $A$ indices of $p$ orbitals correspond to symmetric and antisymmetric combinations with respect to the $z$ axis, i.e.,~$p_i^S = 1/\sqrt2(p_i^t + p_i^b)$, $p_i^A = 1/\sqrt2(p_i^t - p_i^b)$. 
The index $i$ refers to the spatial directions: $x,y,z$ and superscripts $t$ and $b$ indicate the top or bottom chalcogen plane. 
The monolayer is spanned by a hexagonal lattice (see Fig.~\ref{system}) with the spacing $a = \SI{0.319}{nm}$.

The Hamiltonian for each spin component of the system reads,
\begin{equation}
\begin{split}
H =& \sum_{i,\mu\nu} \varepsilon_{i,\mu\nu}^M c_{i,\mu}^\dag c_{i,\nu} + \varepsilon_{i,\mu\nu}^X b_{i,\mu}^\dag b_{i,\nu} \\ 
&+ \sum_{ij,\mu\nu} (t_{ij, \mu\nu}^{MM} c_{i,\mu}^\dag c_{j,\nu} + t_{ij, \mu\nu}^{XX} b_{i,\mu}^\dag b_{j,\nu}) \\ 
&+ \sum_{ij,\mu\nu} t_{ij, \mu\nu}^{MX} c_{i,\mu}^\dag b_{j,\nu} + \text{H.c.},
\end{split}
\label{ham_eq}
\end{equation}
where $i,j$ iterate over lattice sites and $\mu,v$, over atomic orbitals. 
The creation operators $c^\dagger$ and $b^\dagger$ are associated to M and X orbitals, respectively. 
The first term of the Hamiltonian corresponds to the onsite energies with matrix elements
\begin{equation}
\epsilon^M=  
\left(
\begin{array}{ccc}
\epsilon_0  & 0 & 0 \\
0 & \epsilon_2  & -i\lambda_Ms_z \\
 0 & i\lambda_Ms_z & \epsilon_2 \\
\end{array}\right) + \mathbb{1}(s_z E_z -\mu),
\end{equation}
and
\begin{equation}
\epsilon^X=  \left(
\begin{array}{ccc}
\epsilon_p + t_{xx}& -i\frac{\lambda_X}{2}s_z & 0 \\
 i\frac{\lambda_X}{2}s_z & \epsilon_p + t_{yy} & 0 \\
 0 & 0 & \epsilon_z - t_{zz}\\
\end{array}\right) + \mathbb{1}(s_z E_z -\mu).
\end{equation}
where $s_z$ equals 1 $(-1)$ for spin up (down) component.

The second and third sum in Eq.~\eqref{ham_eq} correspond to the hopping elements between intra- and interatomic orbitals, respectively. They are given in Ref.~[\onlinecite{gut_valley_2020}]. 
In the following we take parameters that correspond to \ce{MoS_2} compound~\cite{silva-guillen_electronic_2016, noauthor_notitle_nodate}, but the same model can be applied to other TMDCs as \ce{MoSe_2}, \ce{WS_2}, \ce{WSe_2}. We adopt SO coupling constants $\lambda_M = -\SI{0.086}{eV}$ and $\lambda_S = \SI{0.013}{eV}$, which produce a SO splitting in the conduction band minimum of \SI{3}{meV} and the crossing of the conduction bands, as found in Ref.~[\onlinecite{kormanyos_spin-orbit_2014}]. For the numerical calculations we adopt the ribbon geometry ($L$, $W$) = (200, 10.8) nm. 

We consider a perpendicular magnetic field. The Zeeman splitting is included as $E_{\rm Z}=g\mu_B B/2$, (with Bohr magneton, $\mu_B$, and $g$ factor $g=2$). 
The orbital effects of the magnetic field are incorporated using the Peierls substitution of the hopping elements $t_{nm} \rightarrow t_{nm}\exp\left[-ie\int\bm Ad\bm l/\hbar\right]$ with the vector potential in the Lorentz gauge $\bm A = (-yB,0,0)$. The range of the applied magnetic field is bounded by the critical magnetic field of the superconductors, however, as already shown, high magnetic fields are achievable in planar Josephson junctions~\cite{seredinski_quantum_2019}. We calculate the scattering matrix of the normal region using the \textsc{Kwant} package for quantum transport simulations~\cite{groth_kwant_2014} at $T=0$.

\subsection{Andreev bound states and supercurrent calculation}
\label{sec:abs_calc}
In a SNS junction, the particles and holes in the normal region are Andreev reflected from the superconducting leads when their energy lies within the superconducting gap $|E|<\Delta$. 
In the semiclassical limit, the reflected electrons and holes form periodic trajectories, giving rise to bound states within the superconducting gap $\Delta$, when~\cite{beenakker_universal_1991}:
\begin{equation}
S_{A}(E)S_{N}(E)\Psi_{\mathrm{in}} = \Psi_{\mathrm{in}},
\label{SASN}
\end{equation}
where $\Psi_{\mathrm{in}}=(\Psi^e_{\mathrm{in}}, \Psi^h_{\mathrm{in}})$ describes a wave incident in the junction with electron (e) and hole (h) components and $S_A$ ($S_N$) is the scattering matrix describing Andreev reflections (scattering in the normal part of the junction).

The pairing potential $\Delta$ vanishes in the normal region and therefore taking the hole modes as particle-hole counterparts of the electron modes we can write the scattering matrix of the normal part as
\begin{equation}
S_{N}(E)=\left( \begin{array}{cc}
S(E) & 0 \\
0 & S^*(-E) \\
\end{array} \right),
\label{SN_array}
\end{equation}
which is block diagonal in the electron-hole space and where $S(E)$ describes electronic scattering properties. At the leads, Andreev reflection couples the electron and hole modes and hence the Andreev scattering matrix $S_{A}$ is off diagonal
\begin{equation}
S_{A}(E)=\alpha(E)\left(\begin{array}{cc}
0 & r_A^* \\
r_A & 0 \\
\end{array} \right),
\label{SA_array}
\end{equation}
where $\alpha(E) = \sqrt{1-E^2/\Delta^2}+iE/\Delta$ is the phase factor resulting from matching the wave functions at the normal-superconductor interface.

The Andreev reflection matrix is written in the basis where the outgoing modes are time-reversed partners of the incoming modes,
\begin{equation}
r_{A}=\left(\begin{array}{cc}
i\mathbb{1} & 0 \\
0 & ie^{-i\phi}\mathbb{1} \\
\end{array} \right),
\label{r_array}
\end{equation}
and $\phi$ is the superconducting phase difference. The Andreev reflection process does not mix the modes in the ribbon, which is accounted by the presence of the identity matrix $\mathbb{1}$.

We assume the short-junction limit, when the superconducting coherence length is much larger than the normal channel length $\xi=\hbar v/\Delta\gg L$, with $v$ the Fermi velocity of the modes. This allows us to approximate $S(E)\simeq S(-E) \simeq S(E=0) \equiv s$. Substituting Eqs.~(\ref{SN_array}) and~(\ref{SA_array}) into Eq.~\eqref{SASN}, we obtain the eigenproblem for $\alpha$:
\begin{equation}
\left(\begin{array}{cc}
s^\dagger & 0 \\
0 & s^T \\
\end{array} \right)
\left(\begin{array}{cc}
0 & r^*_{A} \\
r_{A} & 0 \\
\end{array} \right) \Psi_{\mathrm{in}}  = \alpha \Psi_{\mathrm{in}},
\label{ABS_eigenequation}
\end{equation}
whose solution yields the discrete set of Andreev levels with energies $E_i$ [\onlinecite{van_heck_single_2014}].

The complete set of the ABS energies determine the supercurrent through the junction:
\begin{equation}\label{tot_current}
I=-\frac{e}{\hbar}\sum_{E_i>0}\tanh\left(\frac{E_i}{2k_{\rm B}T}\right)\frac{d E_i}{d \phi}.
\end{equation}
Note that we do not assume spin degeneracy for modes and therefore there is no overall factor 2 in the current expression.

Numerically the supercurrent is efficiently calculated following the procedure developed in Ref.~\onlinecite{irfan_geometric_2018}.
Equation~\eqref{ABS_eigenequation} is equivalently written as
\begin{equation}
\left(\begin{array}{cc}
0 & -iA^\dagger \\
iA & 0 \\
\end{array} \right)\Psi_{\mathrm{in}} = \frac{E}{\Delta}\Psi_{\mathrm{in}},
\label{A_eigenequation}
\end{equation}
with 
\begin{equation}
A \equiv \frac{1}{2}(r_As-s^Tr_A).
\label{A_equation}
\end{equation}
Squaring the above equation leads to an eigenproblem for the ABS energies
\begin{equation}
A^\dagger A\Psi_{\rm in}^e=\frac{E^2}{\Delta^2}\Psi^e_{\rm in}.
\end{equation}
Therefore the ABS energy variation with the phase difference $\phi$ in Eq.~\eqref{tot_current} follows readily,
\begin{equation}
\frac{dE_i}{d\phi}= \frac{\Delta^2}{2E_i}\left\langle\Psi_{\mathrm{in}}^e\Big{|}\frac{d(A^\dagger A)}{d\phi}\big{|}\Psi_{\mathrm{in}}^e\right\rangle,
\end{equation}
where $d(A^\dagger A)/d\phi$ is determined analytically from Eqs.~\eqref{r_array} and \eqref{A_equation}. 

The critical current maps are obtained with the use of \textsc{Adaptive} package~\cite{nijholt_textitadaptive_2019}.

\begin{figure*}[ht!]
\includegraphics[width = 14cm]{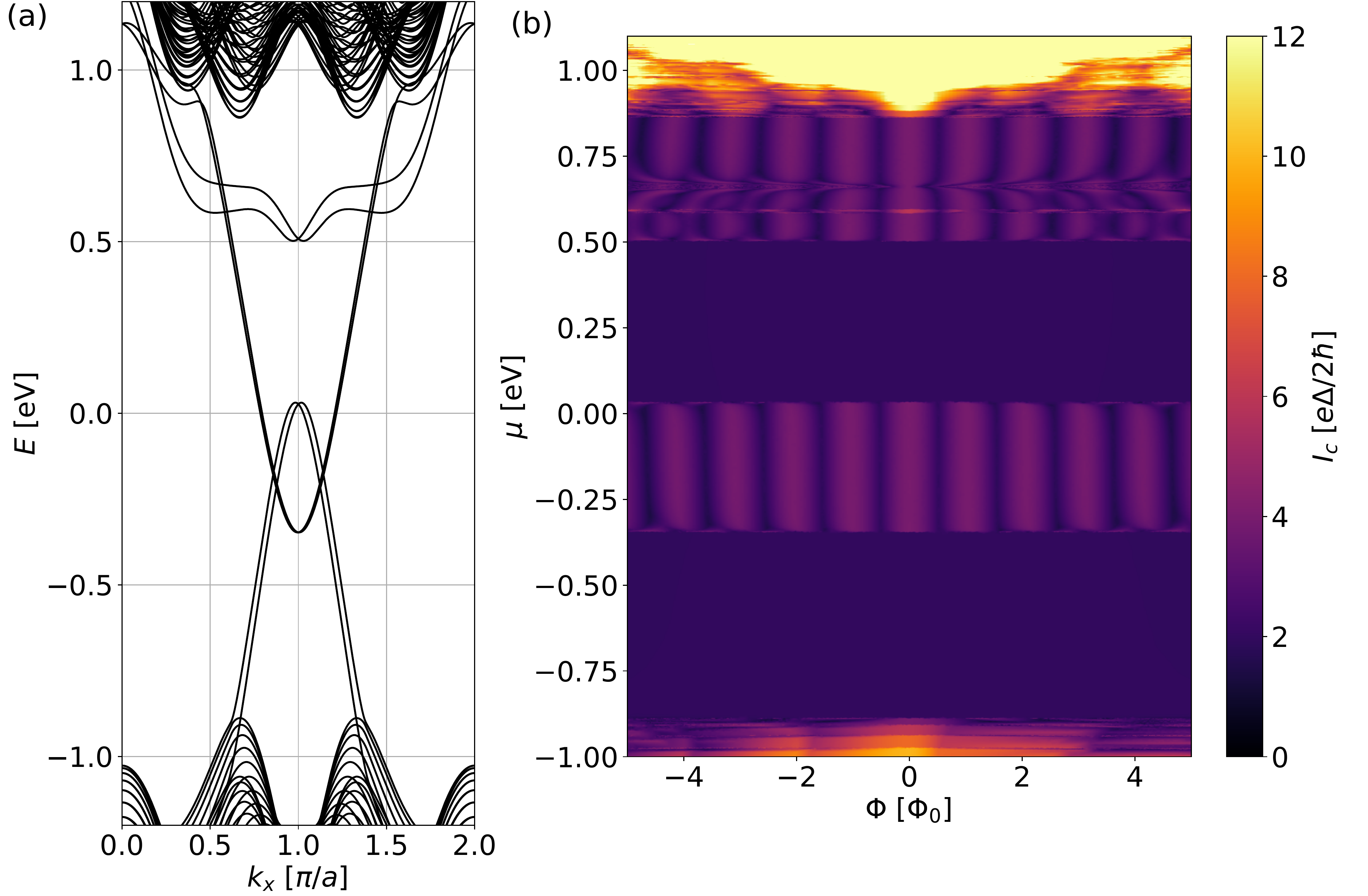}
\caption{(a) Dispersion relation of a \SI{10.8}{nm} wide zigzag \ce{MoS_2} nanoribbon.  (b) Critical current in the Josephson junction embedding the nanoribbon as a function of the external magnetic field and the chemical potential.}
\label{RD_critical_current}
\end{figure*} 

\section{Results}
In Fig.~\ref{RD_critical_current}(a) we plot the dispersion relation of a \SI{10.8}{nm} normal ribbon with zigzag edges in the absence of the magnetic field. In the top and bottom part of the plot we observe a dense sets of bands that correspond to the states in the conduction and valence bands, respectively. In between, there are six bands of the modes located at Mo and S edges of the ribbon. Each edge band comes in a pair of spin-opposite modes split in momentum by strong intrinsic SO coupling that polarizes the spins in a direction perpendicular to the ribbon.

In the map of Fig.~\ref{RD_critical_current}(b) we plot the critical current of the nanoribbon Josephson junction $I_c=\max_{\phi}[I(\phi)]$ as a function of the chemical potential and the magnetic field piercing the system area $LW$ and inducing the flux $\Phi = BLW$. Comparing the critical current pattern with the band structure plotted in Fig.~\ref{RD_critical_current}(a) we see that the character of the supercurrent dependence on the magnetic field is clearly related to the number and type of bands. 
When the chemical potential is set such the Fermi level is crossed only by the bands corresponding to the electrons located on one edge, the critical current is almost constant in $B$.
In contrast, when the Fermi level is crossed by two edge bands the supercurrent exhibits a SQUID-like pattern. Finally, when the chemical potential sets the Fermi level in the bulk spectrum of the conduction band, the current exhibits Fraunhofer-like oscillations.

\subsection{Theory of supercurrent carried by the edge modes}
\label{sec:th_edge}
\begin{figure}[h!]
\includegraphics[width=\columnwidth]{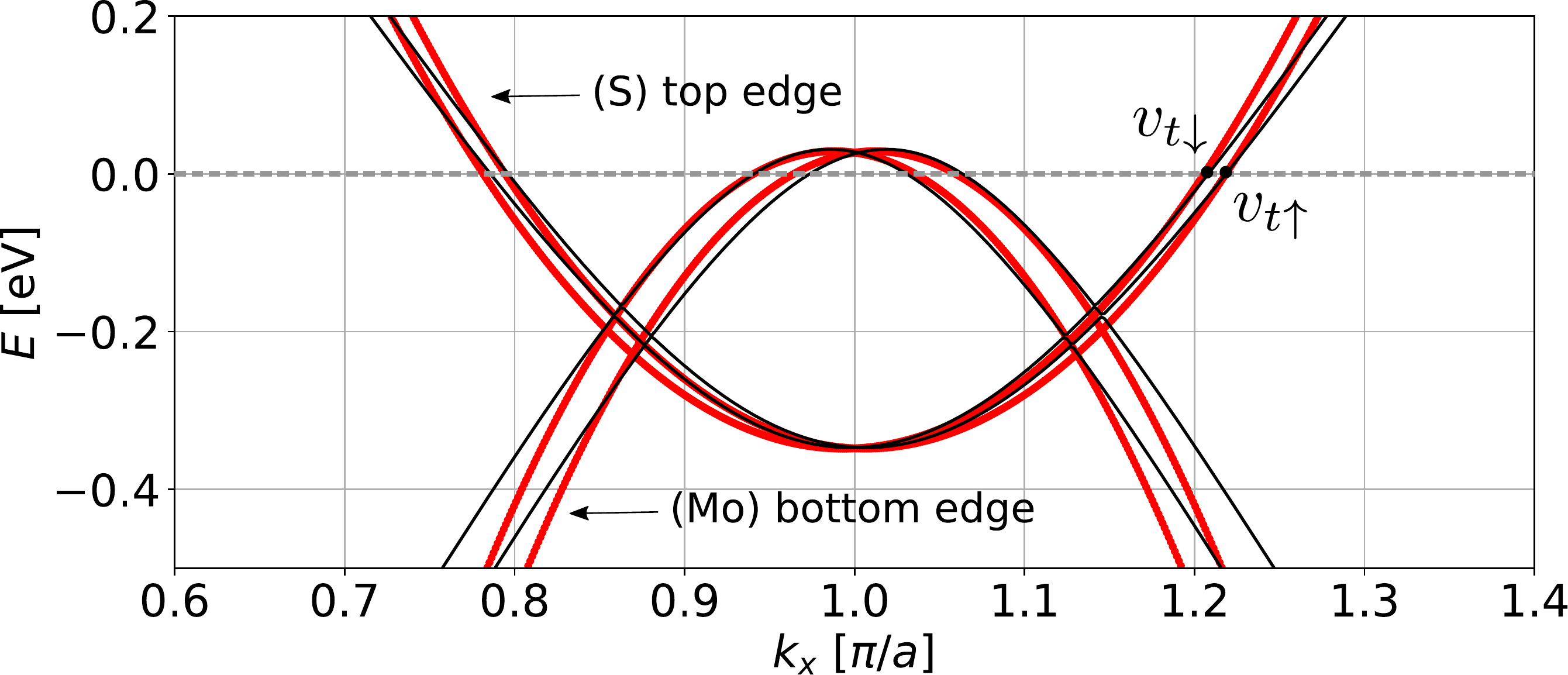}
\caption{The energy dispersion from the tight-binding model Eq.~\eqref{ham_eq} (black curve) and from the one-dimensional continuum approximation of Eq.~\eqref{1D_hamiltonian} (red curves).}
\label{cnt_RD}
\end{figure}

Let us first analyze the ABS energies and the critical current for the chemical potential $\mu$ close to zero. 
In Fig.~\ref{cnt_RD} we present a zoom of the relevant part of the dispersion relation where the bands of the modes located at Mo- (bottom) and S-terminated (top) edges intersect.

\subsubsection{Effective model for edge modes and the ABS spectrum}
We construct an effective one-dimensional model to capture the physics of SO split edge modes under the external magnetic field $B$. The effective Hamiltonian reads
\begin{equation}
H=\left( \begin{array}{cc}
H_t & 0 \\
0 & H_b \\
\end{array} \right) - \mathbb{1}\mu,
\label{1D_hamiltonian}
\end{equation}
which acts on the wave function $\Psi = (\psi_t^\uparrow, \psi_t^\downarrow, \psi_b^\uparrow, \psi_b^\downarrow)$ and where $H_{t,b}$ corresponds (t) top and (b) bottom edge modes, respectively,
\begin{equation}\label{eff_model}
H_{t,b}=\sigma_0\left(\frac{\hbar^2 k_x^2}{2m_{t,b}} - \mu_{t,b} \right) + \sigma_z \alpha_{t,b}k_x + \sigma_z E_{\rm Z},
\end{equation}
with spin Pauli matrices $\bm\sigma$.
The Hamiltonian parameters are the Zeeman energy $E_{\mathrm{Z}}$, the effective masses $m_{t,b}$, band offsets $\mu_{t,b}$,  and the SO amplitudes $\alpha_{t,b}$. The canonical momentum operator in the presence of the magnetic field $B$ reads $k_x=-i\partial_x+eA_x/\hbar$, in a gauge where $\bm A=(-yB,0,0)$.

To find numerically the ABS and supercurrents, we discretize the  Hamiltonian Eq.~\eqref{1D_hamiltonian} on a lattice with spacing $\delta x$. 
The effect of the vector potential is included through the Peierls substitution as a phase on hopping amplitudes $t_{nm}$ between adjacent sites $n$ and $m$,  $t_{nm}^{t,b}\mapsto t_{nm}^{t,b}\exp[ie \delta x W_{t(b)}B/\hbar]$ and $W_{t(b)} = +(-)W/2$ for the mode located at the top (bottom) of the ribbon.
We fit the Hamiltonian parameters to reproduce the bands at $\mu=0$, obtaining $m_t = 0.49\,m_0$, $m_b=-0.3m_0$, $\alpha_t = \SI{10}{meVnm}$, $\alpha_b=\SI{30}{meVnm}$, $\mu_t=\SI{348}{meV}$, $\mu_b=\SI{-27}{meV}$, with $m_0$, the electron rest mass.
The dispersion relation of the full tight-binding and discretized continuum model is shown with black and red curves in Fig.~\ref{cnt_RD}.

The continuum model admits analytical solutions for the ABS in the short-junction limit. 
Following Ref.~\onlinecite{mironov_double_2015}, the positive ABS energies of Eq.~\eqref{1D_hamiltonian} are
\begin{equation}
E_{s\sigma}
=\Delta\left|\cos\left(\frac{\phi}{2}-\frac{seBLW}{2\hbar}
+\frac{\sigma E_{Z}L}{\hbar v_{s\sigma}}\right)\right|,
\label{ABS_analytics}
\end{equation}
where $s=+$ or $t$ ($s=-$ or $b$) for top (bottom) edge, and the spin index $\sigma=+$ ($\sigma=-$) for spin $\uparrow (\downarrow)$ of right-moving modes.
The positive Fermi velocities $v_{s\sigma}$ are evaluated at zero magnetic field and are in general different for the top or bottom right-moving edge modes
\begin{equation}\label{Fermi_velocity}
v_{s\sigma} = \sqrt{2(\mu_s+\mu)/m_s + \alpha_s^2/\hbar^2}.
\end{equation}
The independence of Fermi velocities on the spin $\sigma$ is a peculiarity of the parabolic spectrum, therefore we denote in this section $v_s\equiv v_{s\sigma}$.
As we show in next sections, away from the energy windows near $\mu=0$, where the edge states no longer have a parabolic dispersion, the Fermi velocity of right-moving modes depends on spin projection and is determined numerically. 

\begin{figure}[t]
\includegraphics[width=\columnwidth]{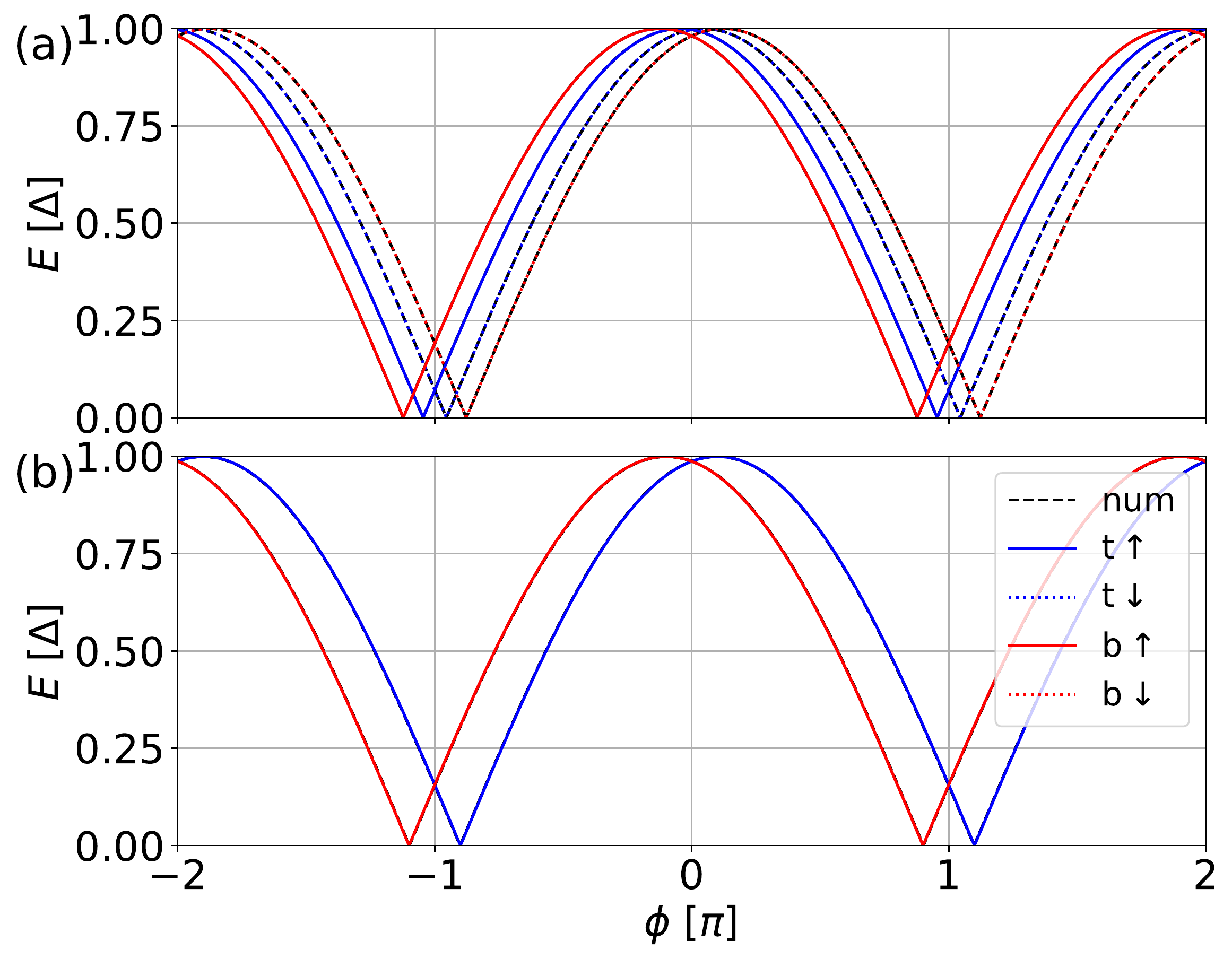}
\caption{ABS spectrum at $\mu=0$ for top (t) and bottom (b) edge modes of spin projection up ($\uparrow$) and down ($\downarrow$). (a) shows only Zeeman interaction effects, which spin-split the ABS. (b) shows the effects due to orbital effects alone. Numerical results (dashed line, ``num'') from the discretized continuum model follow perfectly the analytical dispersion.
The magnetic flux is $\Phi/\Phi_0=2.1$. The panels share the legend.
}
\label{fig:cont_abs}
\end{figure}

The ABS energies Eq.~\eqref{ABS_analytics} depend on the superconducting phase difference ($\phi$), shifted by the magnetic field through orbital and, respectively, Zeeman effects. 
Note that the orbital effects produce shifts proportional to the normal system area $LW$ pierced by the magnetic field $eBLW/\hbar=\pi\Phi/\Phi_0$, where $\Phi_0$ is the magnetic flux quantum.
In contrast, the Zeeman effect produces shifts proportional to the length of the edge channel and lifts the edge degeneracy of the ABS.
To illustrate the two different magnetic field effects, we show in Fig.~\ref{fig:cont_abs} the ABS spectrum in the presence of either Zeeman, or orbital effects.
The analytical solution Eq.~\eqref{ABS_analytics} is also checked against the numerical solutions obtained from the discretized continuum model using the methods of Sec.~\ref{sec:abs_calc}.

\subsubsection{Josephson and critical current}
\begin{figure}[t]
\center
\includegraphics[width = \columnwidth]{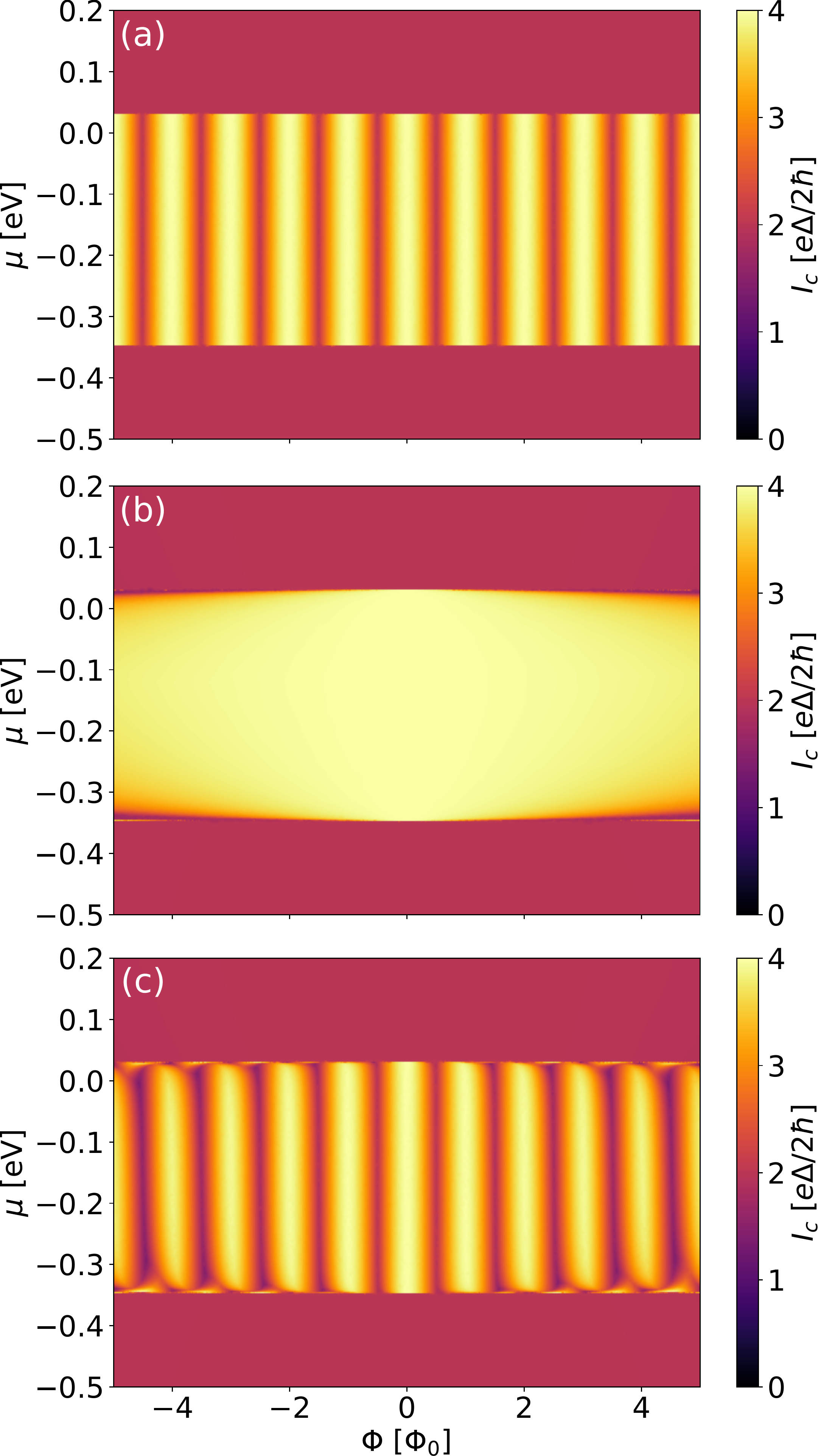}
\caption{Critical current obtained in the effective mass model of the edge states in the presence of (a) only orbital effects, (b) only Zeeman interaction, and (c) both effects.}
\label{cnt_Ic}
\end{figure}

In this section we focus on the Josephson current through the junction and its maximum value, the critical current.
Under the effect of the magnetic field, the critical current exhibits multiperiodic oscillations due to the shifts in the ABS dispersion, induced by the both orbital and Zeeman effects in the junction.~\cite{mironov_double_2015}
To understand separately the two effects, we first calculate numerically the critical currents from the effective model Eq.~\eqref{1D_hamiltonian} for chemical potentials in its region of validity (see Fig.~\ref{cnt_Ic}).

In Fig.~\ref{cnt_Ic}(a), under orbital effects alone, we see that the critical current develops SQUID-like oscillations with period $\Phi_0$ in the region of energetic overlap for the edge dispersion.
Outside that region, the supercurrent is carried by a single edge, and since no magnetic flux is enclosed between the spin up and down modes, the critical current shows no oscillations, and $I_c=e\Delta/\hbar$.

In contrast, under the Zeeman interaction effect alone Fig.~\ref{cnt_Ic}(b), the current displays a slow decay from its maximum value at zero magnetic field. At much higher magnetic fields than shown, $I_c$ displays a beating pattern with long periods in flux.
Noticeably, the critical current varies with the chemical potential, since the ABS depend in this case on $\mu$ through the edge mode Fermi velocities. Outside the overlap region, when only one edge carries the current, one can still observe a slight deviation of $I_c$ from $e\Delta/\hbar$ due the Zeeman interaction effect on the Fermi velocity. 
Finally, the total $I_c$ versus $B$ is shown in Fig.~\ref{cnt_Ic}(c) when both Zeeman and the orbital effects are included.

To get analytic insight into the numerical results, we compute the Josephson current Eq.~\eqref{tot_current} given by positive-energy subgap states from Eq.~\eqref{ABS_analytics},
\begin{eqnarray}\label{curr}
I=-\frac{e}{\hbar}\sum_{s\sigma}
\tanh\big(\frac{E_{s\sigma}}{2k_BT}\big)\frac{dE_{s\sigma}}{d\phi}.
\end{eqnarray}

In the zero-temperature limit of the tight-binding simulations, the current reduces to  
\begin{equation}\label{zero_current}
I= \frac{e\Delta}{2\hbar}\sum_{s\sigma}
\text{sgn}[\cos(x_{s\sigma})]
\sin(x_{s\sigma}),
\end{equation}
where
\begin{equation}
x_{s\sigma} = \frac{\phi}{2}-\frac{s\pi\Phi}{2\Phi_0}+\frac{\sigma E_{Z}L}{\hbar v_{s}}.
\end{equation}
Each of the four right-moving edge modes contributes to carrying a maximum critical current of $e\Delta/2\hbar$.~\cite{beenakker_fermion-parity_2013}

To get better insight into the numerical results, it is useful to Fourier analyze the zero-temperature current,
\begin{eqnarray}
I&=&\frac{e\Delta}{\hbar}\sum_{n=1}^{\infty}
\frac{(-1)^{n+1}}{\pi}\frac{8n}{4n^2-1}\notag\\
&&\times\sum_s \sin\left(n\phi-\frac{sn\pi\Phi}{\Phi_0}\right)
\cos\left(\frac{2n E_{\rm Z}L}{\hbar v_s}\right).
\end{eqnarray}
The first harmonic $n=1$ is dominant and gives the leading behavior of the supercurrent.
The analysis of the first harmonic is also useful since it is directly proportional to the high-temperature current obtained in the limit $k_{\rm B}T > \Delta$,
\begin{equation}\label{ht_current}
I=2I_0\sum_{s}\sin\left(\phi-\frac{s\pi\Phi}{\Phi_0}\right)
\cos\left(\frac{2E_{\rm Z}L}{\hbar v_s}\right),
\end{equation}
with $I_0=e\Delta^2/8\hbar k_{\rm B}T$, the maximal critical current carried by a single spin-resolved edge mode.

In order to make further analytical progress, we will focus in the following on the dominant harmonic of the zero-temperature current, or equivalently, on the high-temperature current.
We investigate the effect of the magnetic field on the critical supercurrent in two limit cases, when only orbital effect is present, and when only the Zeeman effect is present.

\begin{figure}[t]
\includegraphics[width=\columnwidth]{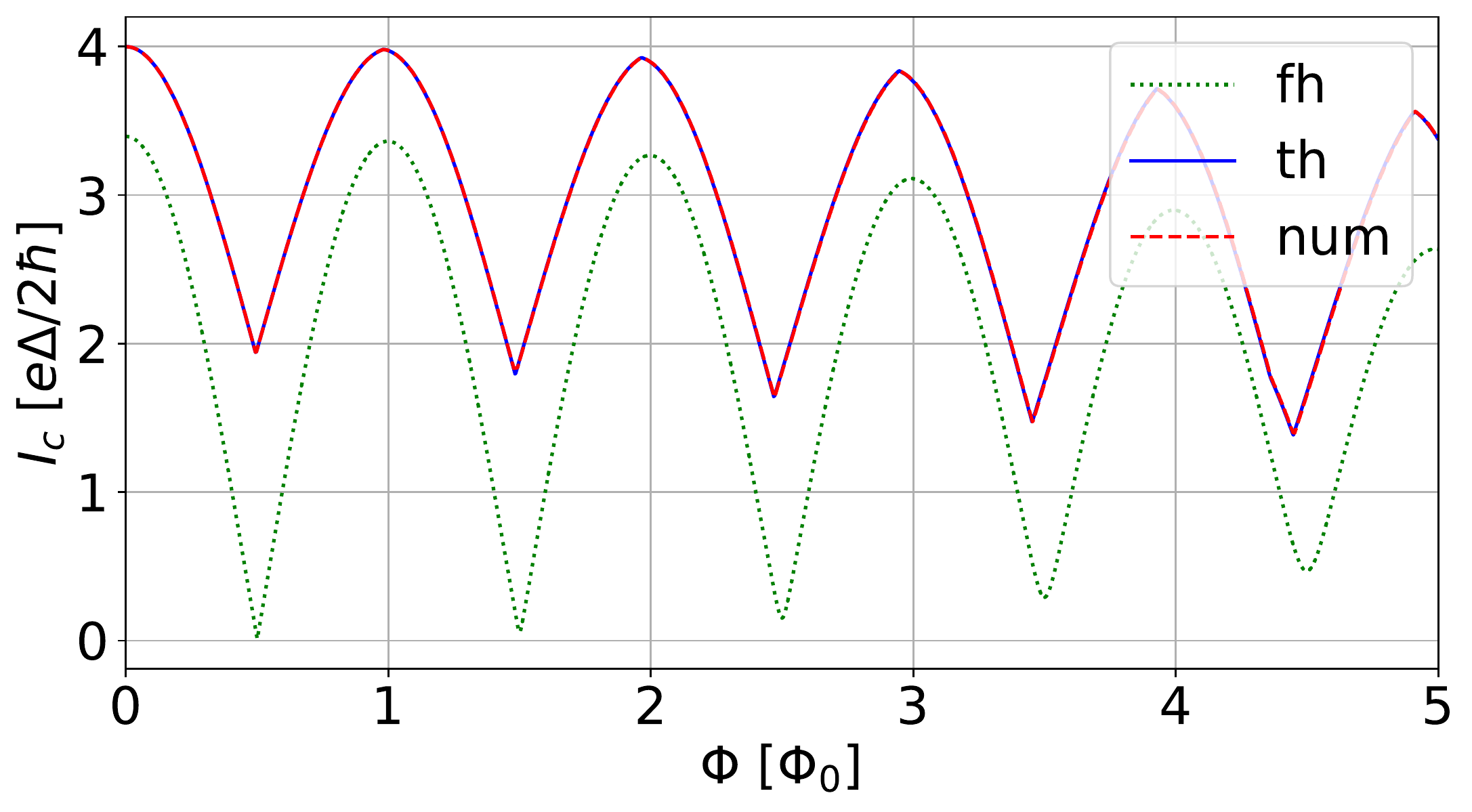}
\caption{Critical current in units of $e\Delta/2\hbar$ at $\mu=0$ as a function of magnetic flux in units of $\Phi_0$. The blue line (th) that represents $I_c$ computed from the analytical solution to continuum model is overlapped with the dashed red line (num) that depicts  $I_c$ computed numerically from the discretized continuum model.
The dotted green line denotes the first harmonic (fh) of the zero-temperature $I_c$, and it is proportional to the high-temperature $I_c$.
}
\label{fig:cross_ic}
\end{figure}

When the Zeeman interaction is absent in the model, and only the orbital effect are present, the Andreev energies are degenerate in spin. 
The critical supercurrent determined from Eq.~\eqref{ht_current} reads
\begin{equation}
I_c = 4I_0\left|\cos\left(\frac{\pi\Phi}{\Phi_0}\right)\right|.
\end{equation}
The critical current has a characteristic SQUID pattern seen in numerics in Fig.~\ref{cnt_Ic}(a), with a period $\Phi_0$.

If only Zeeman effect is present in the model (no orbital magnetic effects), then the critical current obtained from Eq.~\eqref{ht_current} reads
\begin{eqnarray}
I_c &=&2 I_0\left|\sum_{s}\cos\left(\frac{2E_{\rm Z}L}{\hbar v_s}\right)\right|,\\
&=&4I_0\left|\cos\left(\frac{E_{\rm Z}L}{\hbar}
\left(\frac{1}{v_t}+\frac{1}{v_b}\right)\right)
\cos\left(\frac{E_{\rm Z}L}{\hbar}\left(\frac{1}{v_t}-\frac{1}{v_b}
\right)\right)\right|.
\notag
\end{eqnarray} 
The critical current exhibits a beating pattern seen only for very large magnetic fields. 
As expected from numerical solutions shown in Fig.~\ref{cnt_Ic}(b), the critical current depends on the chemical potential in the region of overlap for the edge states through the Fermi velocities $v_{b,t}$.

When both Zeeman and orbital effects are present, we plot in Fig.~\ref{fig:cross_ic} a cross-section at $\mu=0$ of the critical current map from Fig.~\ref{cnt_Ic}. 
The analytical result shown on the same plot captures perfectly the behavior seen in numerics.
Also we plot the first harmonic of the current, which captures only qualitatively the pattern of the full zero-temperature current. 

\begin{figure}[h!]
\center
\includegraphics[width = 7cm]{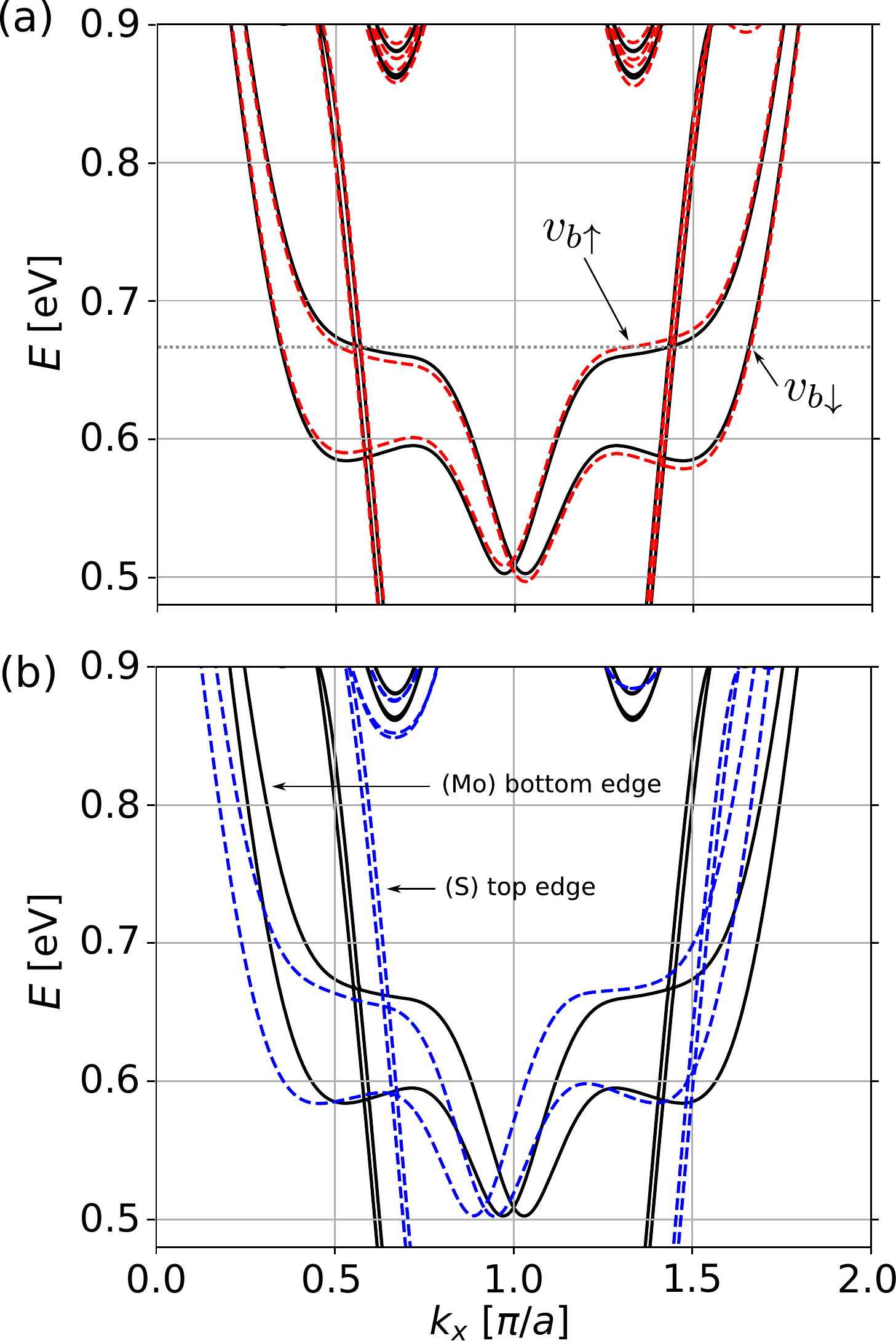}
\caption{Black curves show the dispersion relation of $\mathrm{MoS_2}$ ribbon close to the conduction band. Panel (a) with the red curves shows the bands obtained in the presence of Zeeman interaction while (b) shows bands in the presence of orbital effects of the magnetic field.}
\label{RD_bend_bands}
\end{figure}

\subsection{Anomalous effects due to the presence of strongly non-parabolic bands}
\label{sec:nonpar}
Let us now focus on the energy regime close to the conduction band minimum, see Fig.~\ref{RD_bend_bands}, where the edge modes have a dispersion deviating strongly from the parabolic character.
In this energy window there is again current carried on both top edge, through orbitals localized on S atoms, and bottom edge, through orbitals localized on Mo atoms. In contrast to the case studied in the previous section, the Mo edge band has electron character, and its modes can show a large difference in velocities depending on spin projection. 

\subsubsection{Zeeman interaction}
In Fig.~\ref{RD_bend_bands}(a) we plot zoom-ins on the band structure without (black curves) and with the Zeeman interaction included (red dashed curves). As the edge modes have well-defined spins in $z$ direction, the perpendicular magnetic field increases (decreases) energies of spin up (down) states by $E_{\rm Z}$. 

In the dispersion relation we observe a set of bands corresponding to the states localized on the Mo edge that have strongly non-parabolic character. When the chemical potential is tuned such that the bands cross the Fermi level in the non-parabolic regime, the two spin-opposite modes on one edge will significantly differ in Fermi velocity for each direction of propagation [see the dashed line in Fig.~\ref{RD_bend_bands}(a), $v_{b\uparrow}\neq v_{b\downarrow}$]. 
The different velocities for outer and inner branches of the Mo edge modes will result in unequal phase shifts of the two ABS corresponding to this edge as introduced by the Zeeman term in~Eq. (\ref{ABS_analytics}).


\begin{figure}[h!]
\includegraphics[width = \columnwidth]{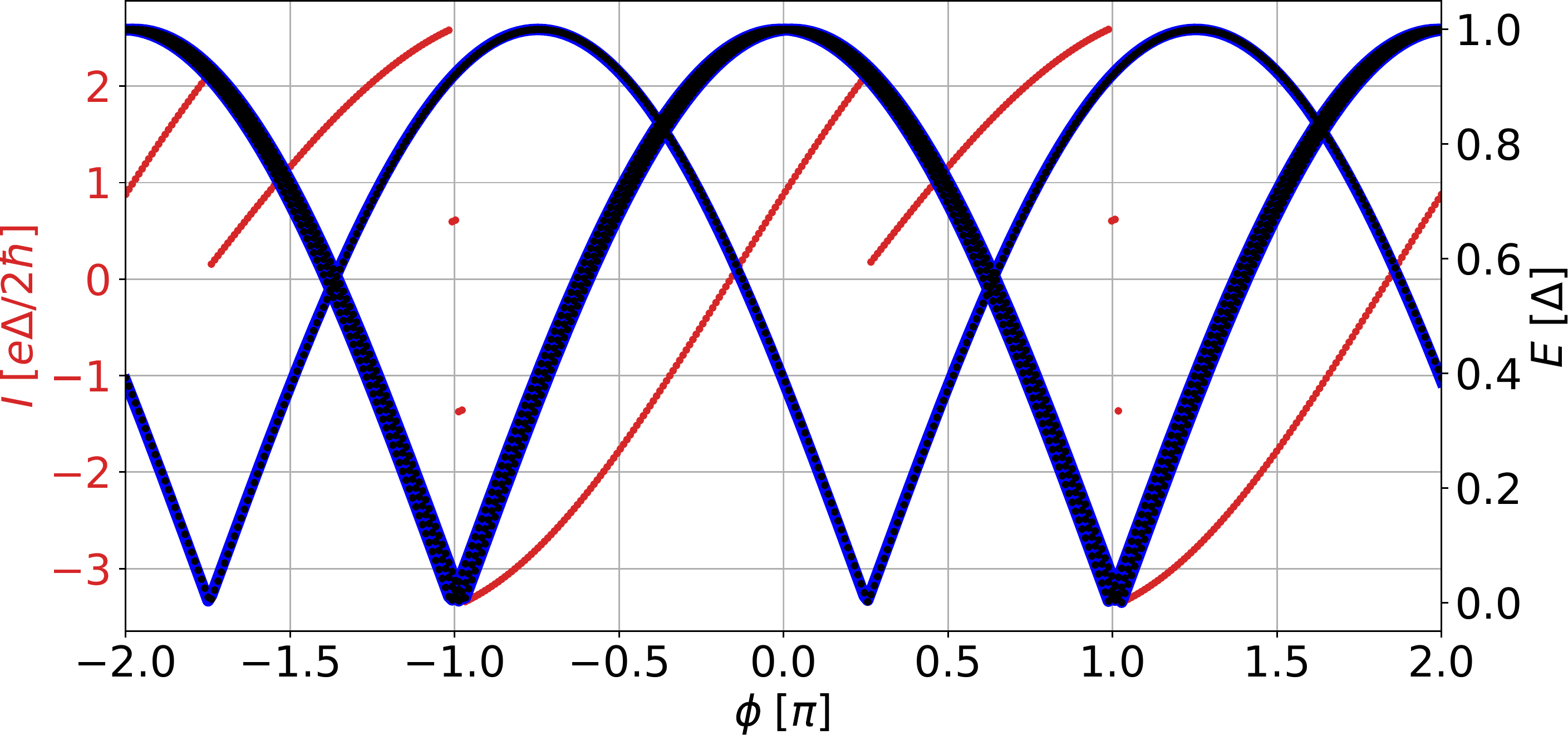}
\caption{Andreev bound states calculated in the tight-binding model (black dots) overlapping the evaluation of Eq.~\eqref{ABS_analytics} (blue curves) and the supercurrent (red) in the presence of Zeeman interaction for $B=\SI{400}{mT}$ and $\mu = \SI{0.66}{eV}$.}
\label{Zeeman_abs}
\end{figure} 

The ABS energies follow from Eq.~\eqref{ABS_analytics}, which remains valid near the Fermi level, since the edge modes have the same geometrical localization, while details of the energy dispersion enter through a modification of the mode Fermi velocity.
The Fermi velocities are extracted numerically from the band structure of Fig.~\ref{RD_bend_bands} at $\mu=\SI{0.66}{eV}$, and the resulting ABS spectrum is plotted in Fig.~\ref{Zeeman_abs} with blue curves.
In Fig.~\ref{Zeeman_abs} we also plot with black dots the ABS spectrum obtained completely numerically using methods of Sec.~\ref{sec:abs_calc}, without any of the above analytical approximations, and find perfect agreement with the theoretical prediction.
Note that due to small $g$ factor in TMDCs, for $B=\SI{400}{mT}$ there is only a single ABS that is shifted in phase by a considerable amount.
The other three (the other ABS of Mo edge and two ABS on S edge) remain to a good approximation insensitive to the Zeeman interaction, which results in an anomalous ABS structure, $E_i(\phi)\neq E_i(-\phi)$.

The general expression for the high-temperature current in the absence of the orbital effects, when all the Fermi velocities are different, reads from Eqs.~\eqref{ABS_analytics} and~\eqref{curr},
\begin{eqnarray}
I&=&2I_0\sum_s\sin\left(\phi +\frac{E_{\rm Z}L}{\hbar}
\left(\frac{1}{v_{s\uparrow}}-\frac{1}{v_{s\downarrow}}\right)\right)\notag\\
&&{}\times\cos\left(\frac{E_{\rm Z}L}{\hbar}\left(\frac{1}{v_{s\uparrow}} 
+\frac{1}{v_{s\downarrow}}\right)\right),
\label{I_different_velocities}
\end{eqnarray}
with $I_0=e\Delta^2/8\hbar k_BT$.
In our case, a further approximation is possible for velocities at $\mu=\SI{0.66}{eV}$, $v_{b\downarrow}\simeq v_{t\uparrow}\simeq v_{t\downarrow}$.
We find that as a result of the anomalous ABS structure, there is finite supercurrent at zero phase difference carried by the non-parabolic band. We plot the supercurrent obtained in the tight-binding calculation with red curves in Fig.~\ref{Zeeman_abs}.
Note that in the numerical calculations we assumed zero temperature. In the experimental scenario, when the temperature is non-zero, we expect that the thermal fluctuations can smooth up the discontinuous jumps in the current. This effect is visible in Fig.~\ref{fig:cross_ic} where the high-temperature current, given by Eq.~(\ref{I_different_velocities}), captures only the highest harmonic of $I_c$ and the jumps are smoothed out at large values of magnetic flux.

\begin{figure}[h!]
\center
\includegraphics[width = 8cm]{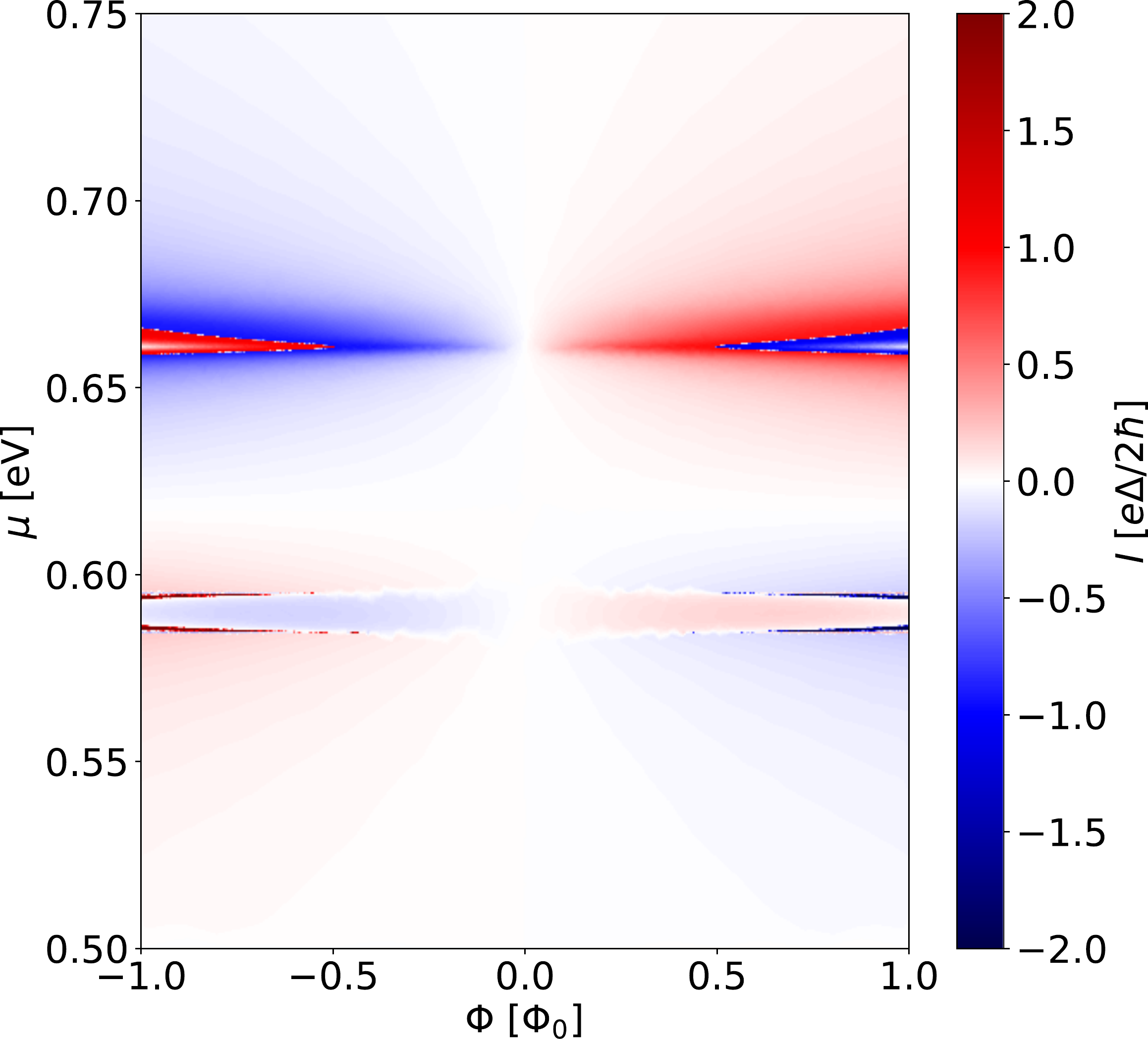}
\caption{Supercurrent at $\phi=0$ carried by the non-parabolic band in the presence of Zeeman interaction as a function of the chemical potential and the magnetic field.}
\label{TB_anomalous_Zeeman}
\end{figure} 

The dependence of the anomalous current carried by the non-parabolic band on the magnetic field and the chemical potential as calculated in the tight-binding model is shown in Fig. \ref{TB_anomalous_Zeeman}. We observe that for the chemical potential values for which the Fermi energy is crossed by non-parabolic bands there is a considerable current for $\phi=0$ present already in a small magnetic field.

\begin{figure}[ht!]
\center
\includegraphics[width = 8cm]{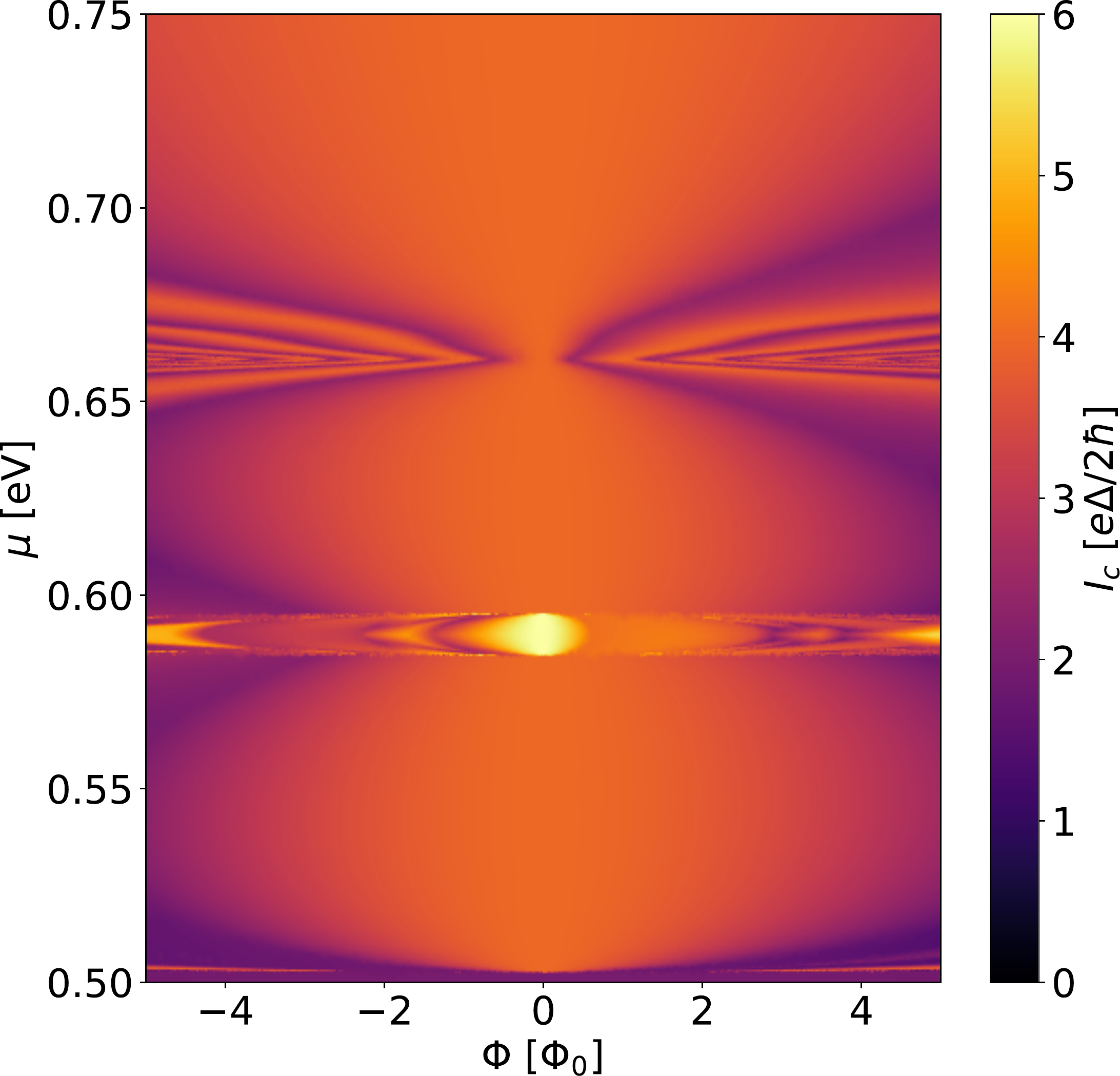}
\caption{Critical current versus the magnetic field and the chemical potential with only Zeeman interaction included.}
\label{critical_current_zoomin}
\end{figure}

Due to the phase shift of the ABS localized on the Mo edge with respect to the remaining three states, the maximal supercurrent in the junction will change when the magnetic field is increased. This effect is demonstrated in Fig. \ref{critical_current_zoomin} where we plot the critical current map as a function of the magnetic field and the chemical potential. We observe a pronounced variation of the critical current whenever the spin opposite Mo bands differ in the Fermi velocity and the anomalous phase shift occurs.

\subsubsection{Orbital effects of the magnetic field}
Let us now focus on the case where the magnetic field is introduced solely through the orbital effects. In Fig. \ref{RD_bend_bands}(b) we observe that the bands corresponding to the opposite edge modes are shifted apart towards opposite values of the wave vector. This results in SQUID oscillations in the critical current. Most importantly, we also find that the orbital effect leads to the valley Zeeman effect that alters the energies of the bulk bands polarized in $K$ and $K'$~\cite{kormanyos_spin-orbit_2014, kormanyos_erratum:_2014}. For a bulk monolayer, the valley splitting due to the orbital part of the magnetic field results from non-zero magnetic moment of the conduction bands~\cite{rostami_valley_2015}. As the modes in the ribbon belonging to the bulk conduction band are also valley polarized we observe band splitting also for the considered wire. Surprisingly, despite the lack of valley polarization of the edge modes~\cite{gut_valley_2020}, we also observe Zeeman-like lifting of their energies induced by the orbital effects. This happens whenever the edge mode wave vector lies in the regime in the dispersion for which the bulk magnetic moment is nonzero, i.e., $k_x \simeq 2\pi/3a, 4\pi/3a$.

\begin{figure}[ht!]
\includegraphics[width = \columnwidth]{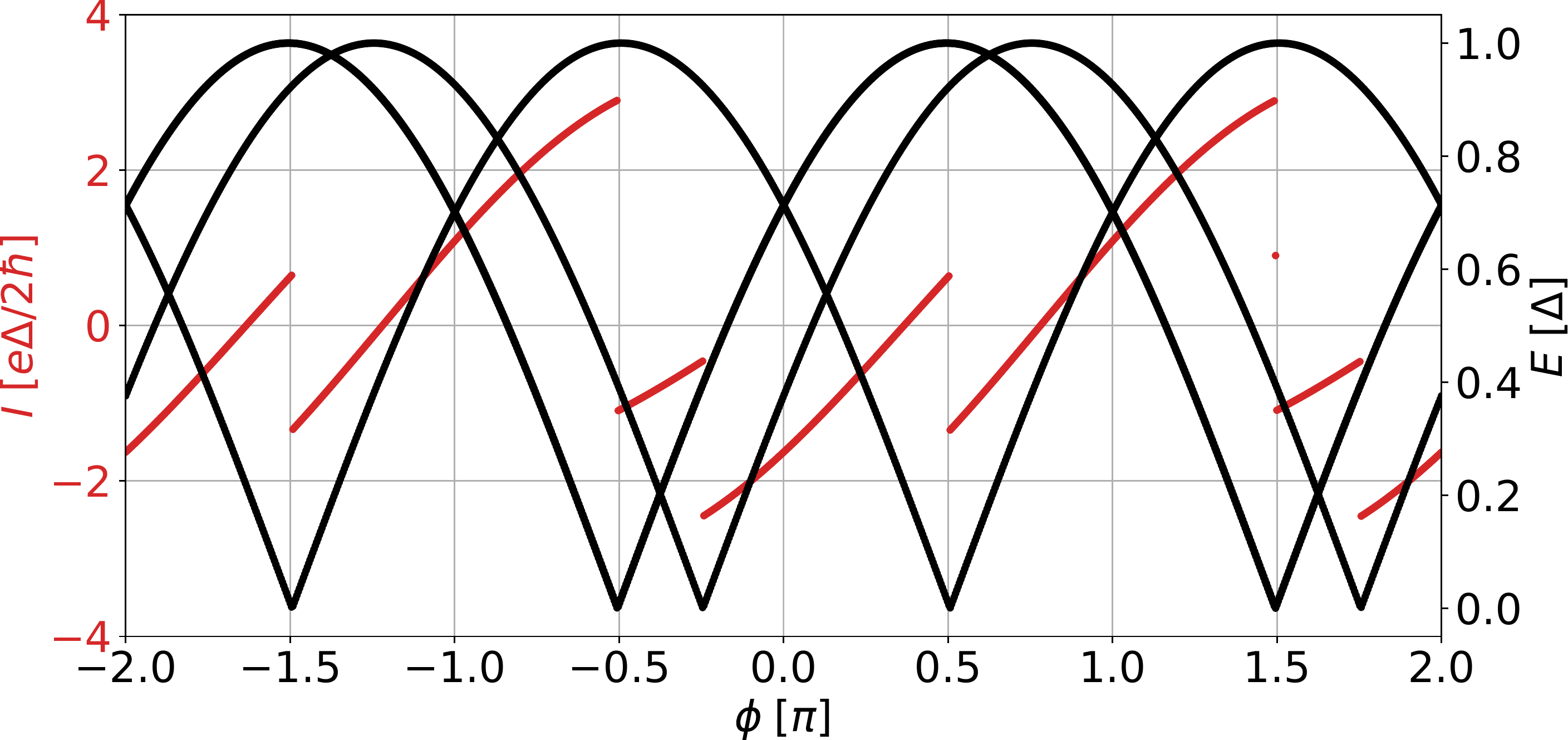}
\caption{ABS spectrum (black) and the supercurrent (red) versus the phase difference between the superconducting leads. 
The results of are obtained for $\mu=\SI{0.66}{eV}$, $B=\SI{500}{mT}$.}
\label{valley_zeeman_abs}
\end{figure}

Taking into account the above mentioned effect, the Eq.~\eqref{ABS_analytics} now reads
\begin{equation}
E_{s\sigma}
=\Delta\left|\cos\left(\frac{\phi}{2}-\frac{seBLW}{2\hbar} + \phi_{V}
+\frac{E_{Vs\sigma}L}{\hbar v_{s\sigma}}\right)\right|.
\label{ABS_analytics_VP}
\end{equation}
The similarity of lifting of the energies of the edge modes by the orbital effects to the ordinary Zeeman splitting is reflected by inclusion of $E_{Vs\sigma}$. This term will act in the same manner as the Zeeman term in Eq. (\ref{ABS_analytics}) and introduce the anomalous shift of non-parabolic bands, see Fig.~\ref{valley_zeeman_abs}.

It is important to note that the gauge choice for the vector potential $\bm A$ is arbitrary as long as the magnetic field $\bm B = \nabla \times \bm A$ does not change. Let us then express the vector potential in a more general form by shifting it with an arbitrary $y'$:
\begin{equation}
\bm A\to \bm A'=[-(y-y')B,0,0].
\end{equation}
Inclusion of the general form of the vector potential into canonical momentum operator results in a wave vector change~\cite{wojcik_durability_2018} of $eBy' /\hbar$ and yields the phase factor for all the ABS $\phi_{V} =  eBLy'/\hbar$ in Eq.~\eqref{ABS_analytics_VP}.

It becomes obvious that, in the presence of the orbital effect, the  specific choice of the vector potential leads to an arbitrary phase shift of the whole ABS structure. This, however, cannot change any of the observables. The anomalous current measurements are performed by putting two Josephson junctions in a loop~\cite{szombati_josephson_2016, mayer_gate_2020} to create a SQUID interferometer, where only the relative shift of the ABS is recorded, and the common, arbitrary phase of the ABS $\phi_{V}$ due to the vector potential is irrelevant.

\begin{figure}[ht!]
\includegraphics[width = \columnwidth]{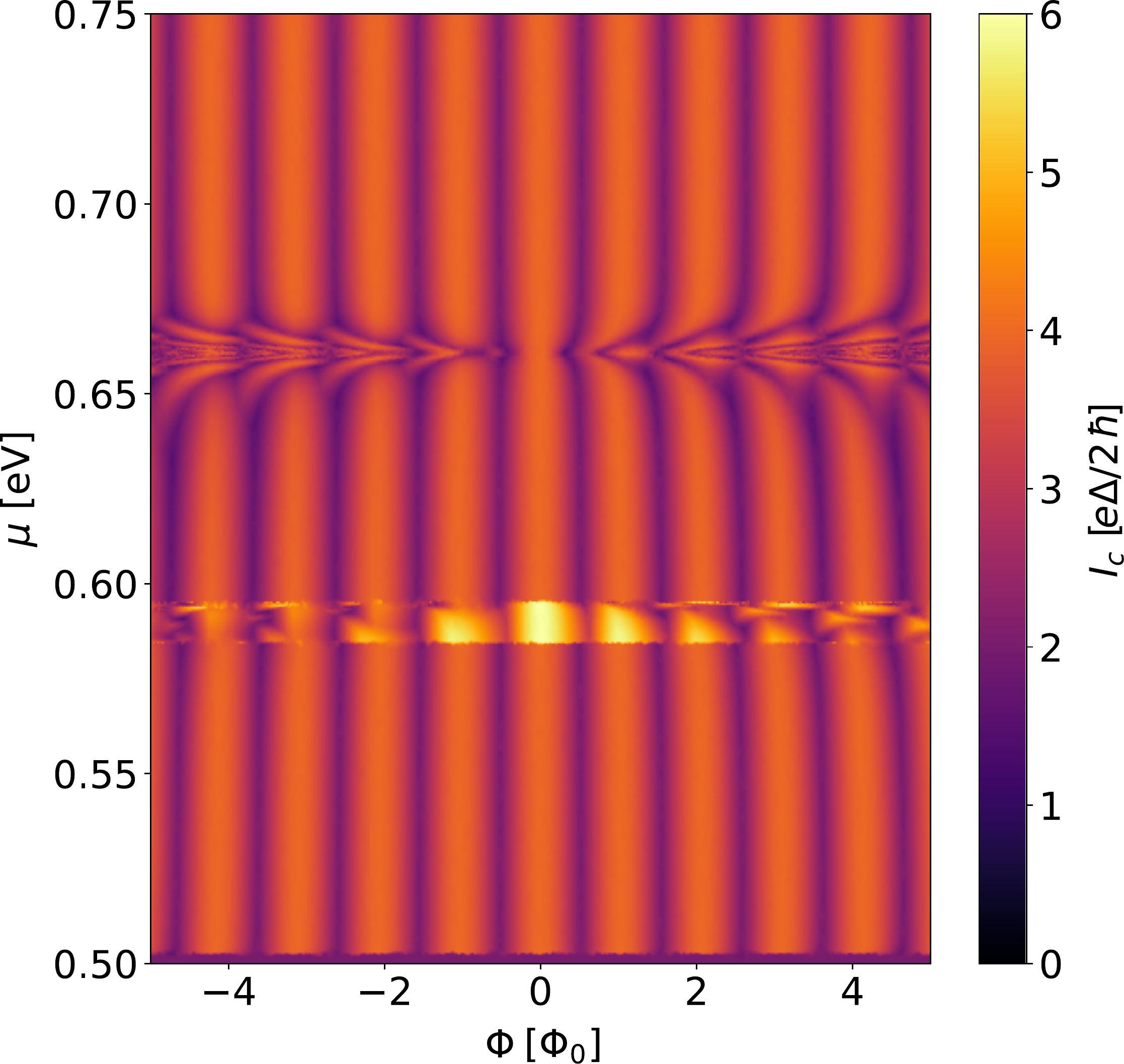}
\caption{Critical current as a function of external magnetic field and chemical potential calculated in the TB model in the presence of only orbital effects of the magnetic field.}
\label{TB_critical_orbital_05075}
\end{figure} 

We take advantage of the fact that a TMDC junction can realize a SQUID interferometer in a single device due to the occupation of the opposite edges of the sample. In the map of Fig.~\ref{TB_critical_orbital_05075} we plot the critical current versus the magnetic field and the chemical potential. We find that the critical current exhibits SQUID oscillations, which are strongly perturbed due to presence of non-parabolic bands. 
Note that since there is no Fermi-velocity-dependent term in the SQUID component of Eq.~\eqref{ABS_analytics_VP}, the deviation from the SQUID pattern results entirely from the Zeeman-like effect that induces the anomalous shift of the ABS due to the presence of non-parabolic edge bands.

\subsubsection{Critical current patterns disclosing the anomalously shifted ABS}
\begin{figure}[ht!]
\includegraphics[width = \columnwidth]{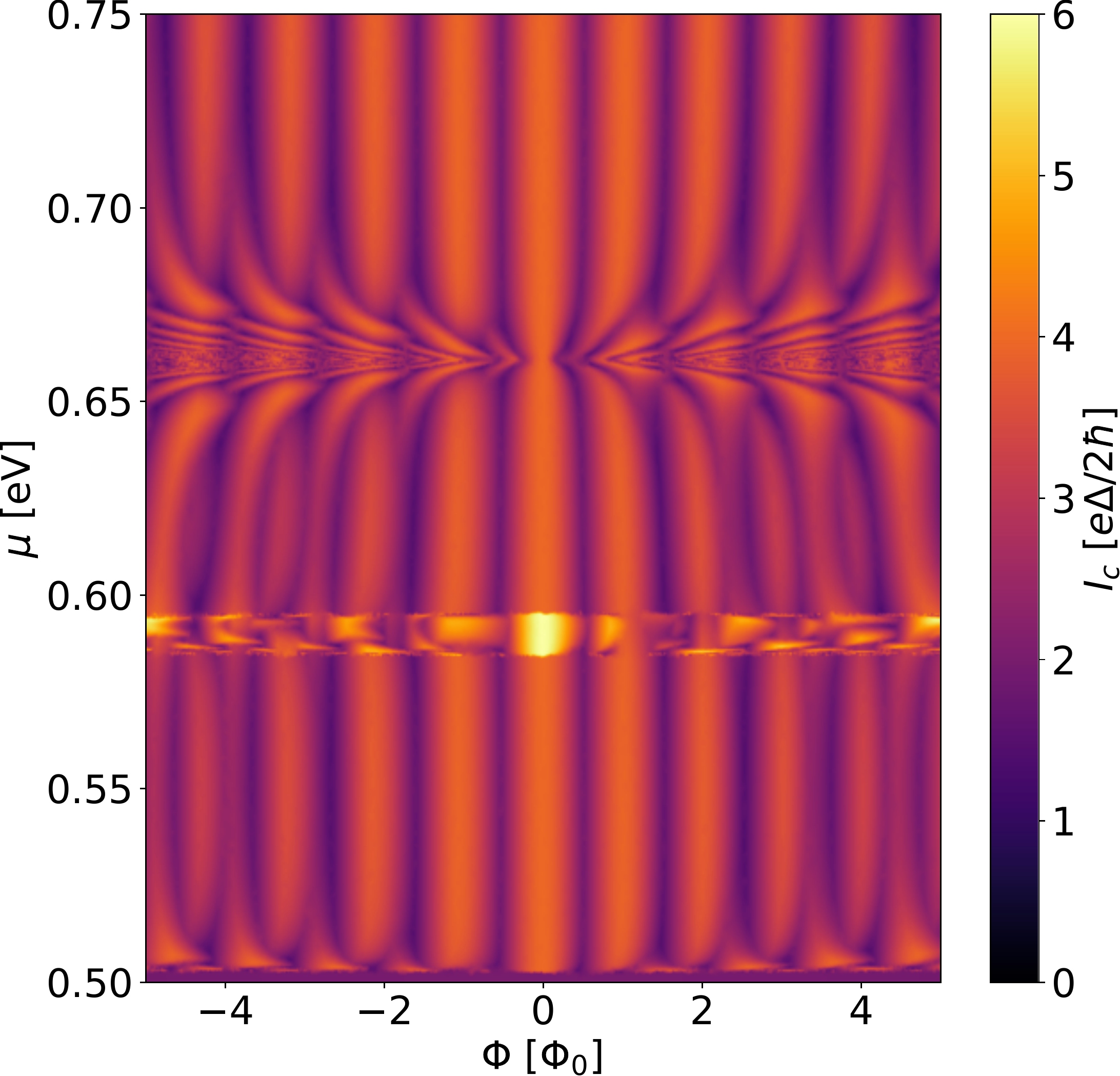}
\caption{Critical current as a function of magnetic flux and chemical potential calculated in the TB model in the presence of both orbital effects and Zeeman interaction.}
\label{TB_critical_orbital_Zeeman_05075}
\end{figure}

We turn our attention to a realistic case when both Zeeman interaction and the orbital effects are present for the magnetic field normal to the TMDC nanoribbon. 
In Fig.~\ref{TB_critical_orbital_Zeeman_05075} we show the critical current versus the field and chemical potential in the junction. 
We clearly see that the strong deviation of the SQUID pattern is a hallmark of the states with a strong non-parabolic dispersion that results in anomalous ABS structure under combined Zeeman and orbital effects of the magnetic field.

\begin{figure}[ht!]
\includegraphics[width = \columnwidth]{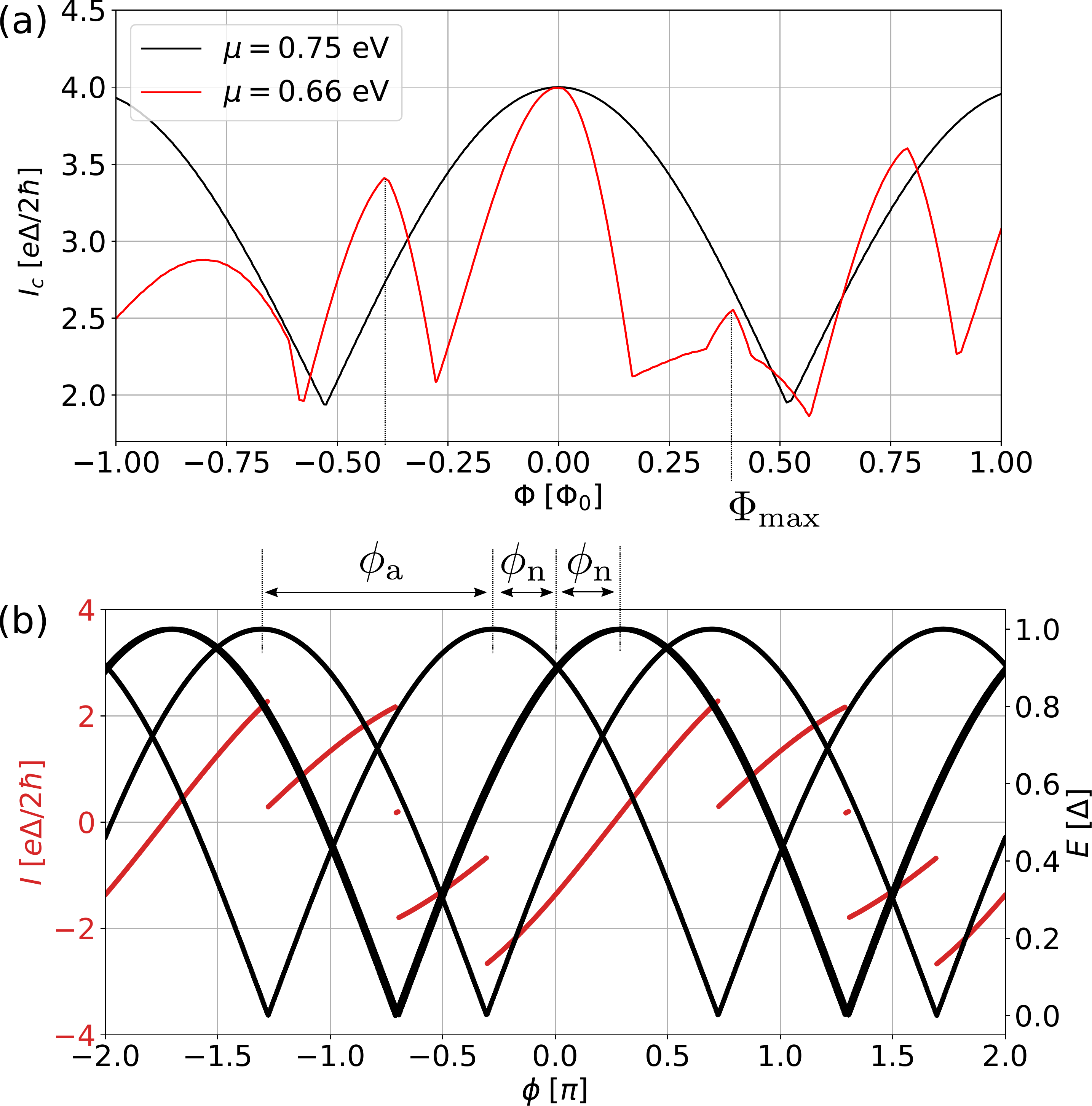}
\caption{(a) Critical current cross section of the map of Fig.~\ref{TB_critical_orbital_Zeeman_05075} for two values of the chemical potential. (b) ABS spectrum for $\Phi=0.81\Phi_{\mathrm{max}}$ and supercurrent versus the superconducting phase difference.}
\label{ic_crossections}
\end{figure} 

Finally, in Fig.~\ref{ic_crossections}(a) we show cross sections of the map Fig.~\ref{TB_critical_orbital_Zeeman_05075} for two values of the chemical potential. For $\mu = \SI{0.75}{eV}$, when there is no anomalous shift, we observe the regular SQUID pattern with the period $\Phi_0$, due to the flux piercing the ribbon.
When $\mu=\SI{0.66}{eV}$ the Fermi energy is crossed by non-parabolic Mo bands and we observe a disruption in the SQUID pattern---new maxima develop at $\Phi_{\mathrm{max}} = \pm039\Phi_0$ [see the black vertical lines in Fig.~\ref{ic_crossections}(a)].

In Fig.~\ref{ic_crossections}(b) we consider the case just before the critical current reaches the first maximum, i.e., $\Phi=0.81\Phi_{\mathrm{max}}$.
We observe that the anomalous ABS corresponding to the Mo modes with smaller velocity is shifted in phase by $\phi_{\mathrm{a}} = (E_{\mathrm{V}b\uparrow}+E_{\mathrm{Z}})L/\hbar v_{b\uparrow}$ from the other Mo ABS which has $\phi_{\mathrm{n}} = eBLW/2\hbar$ shift due to the orbital effects. 
The two other ABS shifted towards positive $\phi$ by $\phi_{\mathrm{n}}$ correspond to the modes located at the S edge. Note that for the estimation of the above phase shifts we neglected the impact of the Zeeman interaction on the normal ABS since its effects is minute [see the two almost degenerate ABS in Fig.~\ref{ic_crossections}(b)].
When the anomalously shifting ABS overlaps with the states located on the opposite edge, i.e., when $\phi_{\mathrm{a}} + \phi_{\mathrm{n}}  = 2 \pi - \phi_{\mathrm{n}}$, the critical current reaches the first maximum and accordingly the anomalous phase shift can be evaluated as $\phi_{\mathrm{a}} = 2\pi - eBLW/\hbar$.

\subsubsection{Impact of the spin-orbit coupling strength}
As obtained from DFT calculations, the value of the spin-orbit gap of a freestanding $\mathrm{MoS_2}$ sheet is $2\Delta=3$ meV \cite{kormanyos_spin-orbit_2014}. Recent experiments, however, suggest that the gap and underlying spin-orbit coupling strength might be sample dependent, as, e.g., observed through Shubnikov-de Haas oscillations in $\mathrm{MoS_2}$ where the gap value was found to be 15 meV \cite{pisoni_interactions_2018}. Therefore, here we inspect the impact of spin-orbit coupling strength on the disruption of the SQUID pattern.

\begin{figure}[ht!]
\includegraphics[width = 0.9\columnwidth]{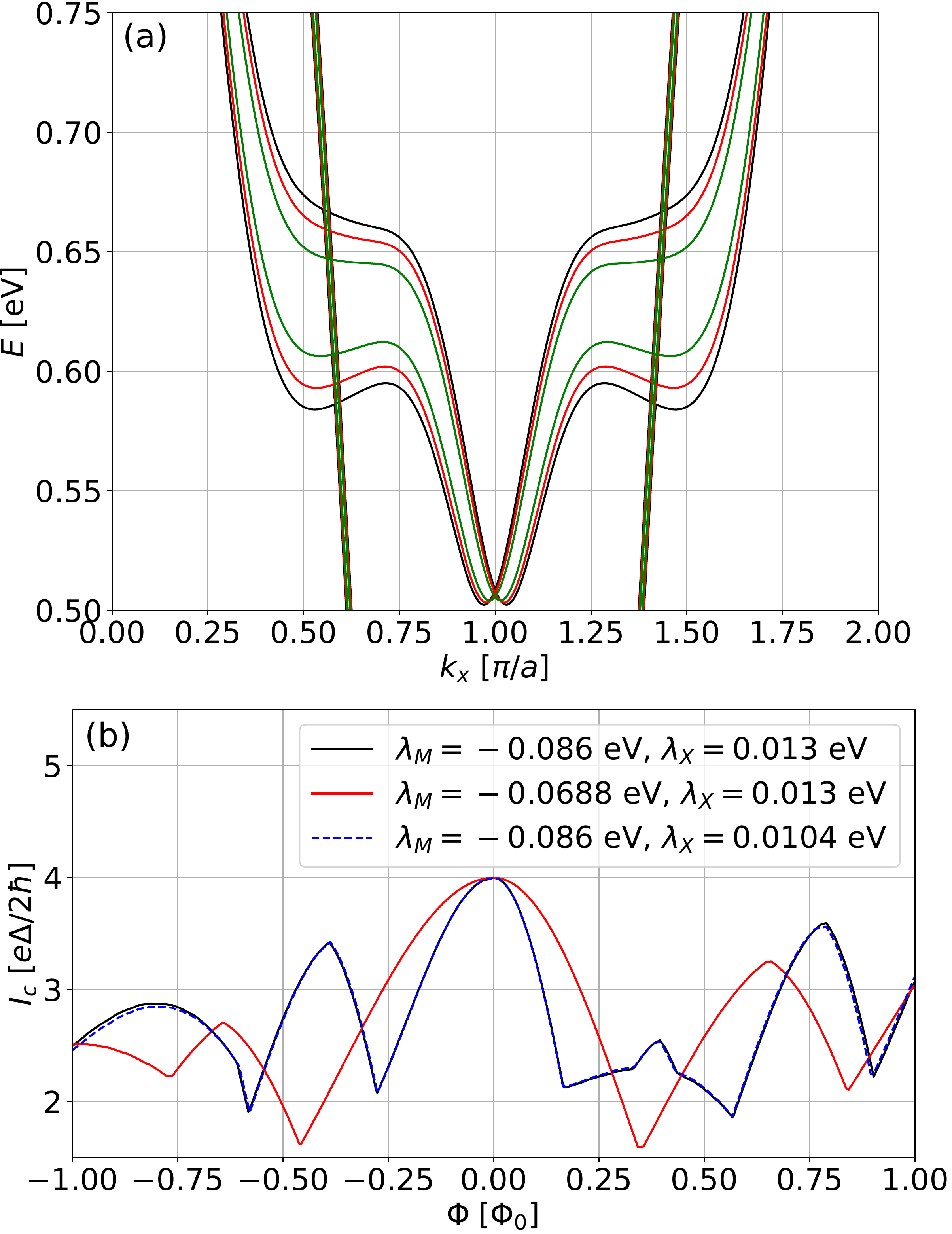}
\caption{(a) Dispersion relation obtained for $\lambda_M =-\SI{0.086}{eV}$, black, $\lambda_M =-\SI{0.0688}{eV}$, red, and $\lambda_M =-\SI{0.043}{eV}$, green. Critical current versus magnetic flux for three values of the spin-orbit coupling parameters.}
\label{varied_SO}
\end{figure} 

In Fig.~\ref{varied_SO}(a) we show dispersion relation of the edge modes for three values of the spin-orbit coupling parameter $\lambda_M$ that controls coupling between the atomic orbitals of Mo atoms. When we lower the absolute value of $\lambda_M$ we observe that the spin-opposite Mo-edge modes decrease their splitting in energy and wave-vector (cf.~black curves obtained for $\lambda_M = -\SI{0.086}{eV}$ with green ones obtained for $\lambda_M = -\SI{0.043}{eV}$). 
As a result, the parts of the dispersion, where a significant difference in Fermi velocities between opposite spin bands occur, become narrower in energy, but the difference itself increases.  Consequently, for small $|\lambda_M|$, the anomalous effect is amplified---in the narrow energy regime where one of the bands becomes flat, while outside of this region the anomalous effect becomes weaker. As a result of the latter, the second SQUID maximum appears for a higher value of the magnetic field, see Fig. \ref{varied_SO} (b).

Finally, we have checked the impact of the strength of spin-orbit coupling of chalcogen orbitals ($\lambda_X$) and found that it has negligible effect on the critical current---cf. black and blue dashed curves in Fig. \ref{varied_SO}(b)---due to small share of occupancy of the S orbitals for the modes located on the Mo-terminated edge.

\section{Summary and conclusions}
\label{sec:summ}
We have studied Josephson junctions formed by a transition metal dichalcogenide nanoribbon placed between two superconducting leads. Using tight-binding model calculations and an analytical approach, we determined the ABS structure and supercurrent in the presence of a perpendicular magnetic field. We explained the separate effects of Zeeman interaction and the magnetic orbital effects on the ABS structure and supercurrent carried by the edge modes of the ribbon. We found that the unusual dispersion relation of the edge modes, with the regimes in which the chiral symmetry is broken, results in the appearance of anomalously shifted ABS in the presence of Zeeman interaction and the orbital effects of the magnetic field. This phenomenon leads to a strong disruption of the SQUID oscillations in the junction when the chemical potential is set such two edges of the ribbon are populated. We point out that this phenomenon can be used experimentally to reveal the presence of non-parabolic edge bands in transition metal dichalcogenide ribbons and to uncover the anomalous phase shift of the ABS in a single SNS junction.

\begin{acknowledgements}
The authors acknowledge fruitful discussions with A. R. Akhmerov, M. Wimmer, T. \"O. Rosdahl, and M. Irfan. M.P.N.~was supported within POIR.04.04.00-00-3FD8/17 project carried out within the HOMING programme of the Foundation for Polish Science co-financed by the European Union under the European Regional Development Fund. D.S.~was supported by CNCS-UEFISCDI, with Project No.~PN-III-P1-1.1-TE-2019-0423. The calculations were performed on PL-Grid Infrastructure.
\end{acknowledgements}
\bibliography{references}

\begin{thebibliography}{60}
\bibitem [{noa()}]{noauthor_notitle_nodate}%
  \BibitemOpen
  \href@noop {} {\bibinfo  {journal} {The parameters of the tight-binding
  non-interacting model are based on DFT simulations where electron-electron
  interactions are considered at a mean-field level}\ }\BibitemShut {NoStop}%
\end{thebibliography}

\end{document}